\documentclass[preprint2]{aastex}

\newcommand{\om}{\Omega_{\rm m}}
\newcommand{\ov}{\Omega_\Lambda}
\newcommand{\ob}{\Omega_{\rm b}}
\newcommand{\oc}{\Omega_{\rm CDM}}

\shorttitle{Evolution of the Cluster Mass Function}
\shortauthors{Bode, Bahcall, Ford, \& Ostriker }

\begin{document}

\title{Evolution of the Cluster Mass Function: Gpc$^3$ Dark Matter Simulations}

\author{Paul Bode, Neta A. Bahcall, Eric B. Ford, and Jeremiah P. Ostriker}
\affil{Princeton University Observatory, Princeton, NJ 08544-1001}

\begin{abstract}
High-resolution N-body simulations 
of four popular Cold Dark Matter cosmologies
(LCDM, OCDM, QCDM, and tilted SCDM), each containing 
$\sim 10^{5}$ clusters 
of galaxies of mass $M_{1.5}>5\times 10^{13}h^{-1}M_{\odot }$
in a Gpc$^{3}$ volume, are used 
to determine the evolution of the cluster mass function from $z=3$ to $z=0$. 
The large volume and high resolution of these simulations allow 
an accurate measure of the evolution of cosmologically important
(but rare) massive clusters at high redshift.
The simulated mass function is presented for cluster masses
within several radii typically used  observationally
($R=$ 0.5, 1.0, and 1.5 $h^-1$Mpc, both comoving and physical)
in order to enable direct comparison with current and future 
observations.
The simulated evolution 
is compared with current observations of massive clusters 
at redshifts $0.3 \lesssim z \lesssim 0.8$.
The $\Omega_m$=1 tilted SCDM model, 
which exhibits very rapid evolution of the cluster
abundance, produces too few clusters at $z\gtrsim 0.3$ and
no massive clusters at $z\gtrsim 0.5$, in stark contradiction with
observations. 
The $\Omega_m$=0.3 models--- LCDM, OCDM, and QCDM--- all
exhibit considerably weaker evolution and
are consistent with current data.
Among these low density models, OCDM evolves 
the least.  These trends are enhanced
at high redshift and can be used to discriminate between
flat and open low density models.

The simulated mass functions are 
compared with  the Press-Schechter approximation. 
Standard Press-Schechter predicts too many low mass clusters
at $z=0$, and too few clusters at higher redshift.
We modify the approximation by a simple parameterization of 
the density contrast 
threshold for collapse, which has a redshift dependence. 
This modified Press-Schechter approximation
provides a good fit to the simulated mass functions.

\end{abstract}


\section{Introduction}  \label{secintro}

The local abundance of clusters of galaxies places a powerful
constraint on 
cosmological parameters: $\sigma _{8}\Omega _{m}^{0.5}$ $\simeq $ 0.5,
where $\sigma _{8}$ is the rms mass fluctuation on an 8$h^{-1}$Mpc scale and 
$\Omega_{m}$ is the present cosmological density parameter (\cite{HA91},
\cite{BC92}, \cite{WEF93}, \cite{ECF96}, \cite{VL96},
\cite{KS97}, \cite{Pen98}). This constraint,
while powerful, is degenerate in $\Omega _{m}$ and $\sigma _{8}$. The
evolution of cluster abundance with redshift, especially for massive
clusters, breaks this degeneracy (e.g., \cite{PDJ89}, 
Oukbir \& Blanchard (1992, 1997),
\cite{ECF96},  \cite{VL96}, \cite{BFC97},
\cite{FBC97}, \cite{CYE97}, 
Henry (1997, 2000), \cite{BF98}, \cite{ECFH98}, \cite{DV99}).
The evolution of high-mass clusters is strong in $\Omega _{m}$ = 1,
low-$\sigma _{8}$ Gaussian models, but is much weaker in low-density, $%
\sigma _{8}$ $\sim$1 models. Therefore, the evolution of cluster
abundance provides one of the most powerful methods to determine both $%
\Omega _{m}$ and $\sigma _{8}$. Various analyses using this method have
shown that the relatively high abundance of observed clusters to 
$z\lesssim 0.8$ indicates that the mass-density of the universe is low: 
$\Omega_{m} \sim 0.2$ - 0.5, and $\sigma _{8} \sim$ 1 (see
references above). However, the total number of clusters currently studied at
high redshifts is still small, and the model comparisons are generally based
on the Press-Schechter (1974) approximation which,  while surprisingly
accurate, is known
to have biases. Direct comparisons with simulations have also been used,
but these
are generally too small to find the rare high-mass clusters, especially
at large redshifts. New observations are currently underway to determine
more precisely the evolution of cluster abundance using optical, X-ray, 
Sunyaev-Zel'dovich,
and gravitational lensing surveys. At the same time, larger and more
accurate cosmological simulations are needed in order to accurately
determine the expected
mass function of clusters of galaxies and its evolution with time,
in order to allow for proper comparison with the upcoming observations. 
%Since massive clusters at high redshifts are very rare,
%large volume simulations are needed in order to reveal the expected
%abundances. 

In this paper we present gigaparsec-scale simulations of
four popular CDM models which allow an accurate determination
of the evolution of the cluster mass function from redshift $z=3$ to
$z=0$, including the rare and cosmologically powerful massive clusters. 
The rapid increase of available computing power and new algorithms
which run efficiently on parallel machines make possible a 
combination of large simulation volume and high resolution.
For comparison, the models presented here are of a larger
volume than the \cite{GBQTBKL99} simulations, and attain 
considerably higher resolution 
than the Virgo ``Hubble Volume'' simulations 
\citep{Evrard98,JFWCCEY00}. 
Furthermore, we are able to cover several different models,
including the first large volume, high resolution simulation
of tilted $\Omega_{m}=1$ as well as a quintessence model.

The aim of this paper is
to provide the foundation for comparisons with the upcoming
observations. 
Thus the evolution of the mass function
is presented in a manner that can be directly compared with observations,
using cluster masses within a variety of specific radii typically
used in observations, rather than using the not 
so easily observable virial mass. 
We discuss the simulations 
and the method of cluster selection in $\S$\ref{seccreate}.
The evolution of the resulting cluster mass function is
presented in $\S$\ref{secevol}.
Comparison with the Press-Schechter
approximation is made in $\S$\ref{secpscomp},
and effects of resolution are discussed in $\S$\ref{secreseff}.
Comparisons with current observations 
are  shown in $\S$\ref{secobscomp}, and conclusions are summarized 
in $\S$\ref{secconcl}.

\section{Creating the Simulated Mass Function}  \label{seccreate}

\subsection{The N-body Simulations}  \label{secsimul}

Four currently favored variants of the CDM model were chosen,
with parameters 
consistent with numerous observational constraints \citep{BOPS99}.  
These observations tend to favor a low density universe with
$\om\simeq0.3$; 
in addition to spatially flat (LCDM) and open (OCDM)
models we include a quintessence model (QCDM).  Like LCDM,
the QCDM is made spatially flat by including a component
with negative pressure, but the Q component is dynamically
evolving and spatially inhomogeneous \citep{CDS98}.  The
equation of state of the Q component, $w=-2/3$,  was chosen to
be in the range favored by observations \citep{WCOS00}.
A standard $\om=1$ CDM model was also run, with a strongly tilted 
power spectrum so as to avoid overproduction of present-day clusters.
All models are normalized (by $\sigma_8$) to match the observed
present-day cluster abundance and be consistent with the
COBE microwave background normalization.
The cosmological parameters for the different runs are listed
in Table~\ref{tblparams}.  

The transfer function for a given model was computed
using the CMBFAST code \citep{SZ96}, modified to handle a
dynamical energy component in the QCDM case \citep{CDS98}.
The resulting power
spectrum was used to generate initial conditions by perturbing
particles on a rectilinear grid; this step was performed with 
the COSMICS software package \citep{MB95,Bert95}.  Each simulation 
was begun at a redshift when the RMS density fluctuation was 15\%.
For QCDM, the fluctuations in the Q component are negligible by
the time the simulation begins ($z\approx 30$). Thus, once the
matter power spectrum has been computed, the Q component 
has an effect only through the overall expansion factor $a(t)$.

All four runs contained $512^3=134$ million
particles in a periodic cube. For the $\om=0.3$ runs the box
length is 1000$ h^{-1}$Mpc; in the SCDM run it is
smaller, such that the volume is reduced by a factor of 0.3
relative to the other models (see Table~\ref{tblparams}).
With this choice of parameters all models, including SCDM, have
an individual particle mass of $6.2\times10^{11}h^{-1}M_\odot$.
Also, the spline kernel softening length in all cases is 
$\epsilon=27 h^{-1}$kpc;  thus there is the same mass and force
resolution in all the models, simplifying comparisons.
The spatial resolution is comfortably smaller than the
$\sim100h^{-1}$kpc characteristic core size of clusters, and
the high mass resolution assures that two body relaxation
is unimportant in the cluster cores.

The N-body evolution was carried out with the TPM code \citep{BOX00}.
The simulations ran on a 128 processor SGI Origin 2000 at
NCSA (as part of the Grand Challenge
Computational Cosmology Partnership); 
a single run took 11 to 14 hours to complete. 
A mesh with 512 cells on a side was used for long-range interactions,
while interactions internal to dense peaks were computed with a tree code.  
The time step for the Particle-Mesh portion of the code is set
such that in a given step 1) the cosmological expansion factor $a$ 
changes by less than 1\%, and 2) no particle being evolved 
solely by the PM code moves more than 1/10 the cell size.
Each tree has its own, possibly smaller, time step.  This was
set such that, for at least 97.5\% of the particles in the tree,
the time step was shorter than 
$0.05*{\rm MAX}(\rho_c^{0.33}, \epsilon)/{\rm MAX}(v, v_{rms})$,
where $\rho_c$ is the density of the cell containing the particle
and $v_{rms}$ is the $rms$ velocity of the entire simulation.
The LCDM run took 469 PM steps from $z=24$ to $z=0$,  and the most
massive tree (which contains one of the denser peaks) took 1887 steps. 

In the TPM code, 
only cells above a given density threshold $\rho_{\rm thr}$
are treated at full resolution, so care must be taken to deal only
with objects which lie above this limit.
The density threshold parameter was set to 
$\rho_{\rm thr}=0.9\bar{\rho}+4.0\sigma$, where $\sigma$ is the 
dispersion of the cell densities;  since $\sigma$ rises over time,
$\rho_{\rm thr}$ is initially only slightly larger than $\bar{\rho}$
and becomes larger as structure forms
(see \cite{BOX00}).  The evolution of $\rho_{\rm thr}$ 
is similar in the three $\om$=0.3 models, rising from $3\bar{\rho}$ at
$z$=10 to $10\bar{\rho}$ at $z$=1;  at $z$=0, $\rho_{\rm thr}=17\bar{\rho}$
for LCDM and slightly less for the other two models.  Since structure
formation happens later in the SCDM model, $\rho_{\rm thr}$ is much
lower, rising from $4\bar{\rho}$ at $z$=1 to $11\bar{\rho}$ at $z$=0.
In this paper we consider collapsed objects above 
$5\times10^{13}h^{-1}M_\odot$, or 80 particles within a radius smaller
than the code cell size, which implies a density well above 
$\rho_{\rm thr}$ at all times. In the LCDM run, the smallest object 
picked out for full resolution treatment contained 28 particles at
$z$=0.5 and 42 particles at $z=0$; the other runs located even
smaller objects.  Thus we are confident that resolution is not a
problem for the objects we are considering.
This will be discussed further in $\S$\ref{secreseff}.

The present models are the largest high-resolution N-body
simulations of OCDM, QCDM,
and tilted SCDM models currently available.
Compared to the Virgo ``Hubble Volume'' simulations 
of LCDM and $\tau$CDM \citep{Evrard98,JFWCCEY00}, 
the simulated box size $L_{\rm box}$  of these models is
a third smaller, 
but mass and force resolution is superior by a factor of three.
\cite{GBQTBKL99} have carried out an OCDM model with
cosmological
parameters similar to the run presented here and mass resolution
a factor of three higher, but with $L_{\rm box}=500h^{-1}$Mpc.
In $\S$\ref{secreseff} 
we will discuss
one additional higher resolution run of the LCDM model, which uses
more than $10^9$ particles.

\subsection{Cluster Selection}  \label{secselect}

Once a simulation is complete, it is necessary to identify
collapsed objects; 
this section describes the algorithm used to identify clusters.
For a given cluster we are interested in locating the total mass 
within a set radius $R_c\sim 1 h^{-1}$Mpc, 
either comoving or physical.

We begin by creating a list of possible cluster centers,
using the HOP package developed by  \cite{EH98}.  
This code calculates a density for each particle, and
associates particles with nearby neighbors possessing a
higher density; thus a particle at a density maximum
becomes the nucleus of a group.
HOP tends to break up larger clusters into a number of smaller groups;
we do not attempt to reunite these with the REGROUP portion of the
HOP package but instead all group positions are treated as 
potential cluster centers.
However, groups with less than 25 particles or  with central density
$\rho_c<180$ are eliminated, because including them in the
analysis only results in more poor 
clusters with mass less than $2\times10^{13}h^{-1}M_\odot$, well
below our mass threshold.

Nearby pairs of clusters are merged together, working out
to a separation of 1$h^{-1}$ Mpc;  the position of a merged cluster 
is set to that of the denser of the two progenitors.
After this merging is complete, the position of each
cluster is revised to the center of mass of all particles
located within 1$h^{-1}$ Mpc of the original position.
Using this final set of cluster positions, 
all particles within a specified radius $R_c$ of
a cluster center are assigned to that cluster.  If a particle is within 
$R_c$ of multiple clusters, it is assigned to the group to
which it has a greater binding energy (using the previously computed
mass and central position of each group).

Note that these cluster masses, defined specifically within
a fixed radius $R_c$ (for example $R_c=1.5h^{-1}$Mpc) are not
the same as virial cluster masses. 
For a rich cluster with $R_{vir}>1.5h^{-1}$Mpc
(where $R_{vir}$ is the radius containing a mean density of 178
times the critical value), 
$M_{1.5}$ will be smaller than the virial mass and
for poor systems $M_{1.5}$ will be larger than the virial mass
when $R_{vir}<1.5h^{-1}$Mpc.
Using a fixed $R_c$ facilitates a proper comparison with
observations, since the virial radius cannot easily be
determined observationally.

\section{Evolution of the Cluster Mass Function} \label{secevol}

Using the selection criteria of $\S$\ref{secselect},
the cluster mass function (MF), which represents the number density
of clusters above a given mass threshold, is determined from the simulations
as a function of redshift for $z=$ 0, 0.17, 0.5, 1, 2, and 3. 
At $z=0$, approximately $10^5$ clusters with
$M_{1.5}>5\times 10^{13}h^{-1}M_\odot$ are located within the
simulation volume.
The MF is determined for cluster masses within different radii--- 
$R_c=$ 0.5, 1, and 1.5$h^{-1}$Mpc, both comoving and 
physical--- in order to explore the
dependence of the MF on this scale, as well as provide for proper
comparisons with different observations.

The cluster MF is presented in 
Figure~\ref{figmfevol} for the four models
studied. The cluster mass used here is the mass within the `standard' 
1.5$h^{-1}$Mpc comoving radius, 
frequently used in observations.
At $z=0$, all models yield the same MF,
designed to be consistent with observations; this consistency is 
essentially forced by
the selected $\sigma_{8}$ normalization of each model (Bahcall and Cen
1992, White  et al. 1993, Eke  et al. 1996, Pen 1998). The
models are thus degenerate (by construction) at $z\sim$0,
at least for the richer clusters (SCDM shows many more poor clusters).
At higher redshifts, a
negative evolution of the cluster MF is seen in all models, as expected,
reflecting the growth of clusters with time from small density fluctuations
(e.g., Press \& Schechter 1974, Peebles 1993, Eke  et al. 1996, Fan 
 et al. 1997). 
 
The \emph{rate} of the evolution, however, is strongly
model dependent; it breaks the degeneracy seen at  $z\simeq 0$. 
The $\Omega_{m}$=1 tilted SCDM model, with much
stronger evolution, predicts orders-of-magnitude fewer clusters at high
redshift than do the low-density models. For example, 
for a given mass the abundance of
clusters in SCDM at $z=0.5$ is comparable to the abundance at $z=1$ in the
low-density models.
Similarly, the number
of $z=1$ clusters in SCDM is comparable to the number of $z=2$ clusters in
the low-density models. At $z=0.5$, the abundance of Coma-like clusters
(with $M_{\rm 1.5com}\simeq 6\times 10^{14}h^{-1}M_{\odot }$) 
is 100 times larger in
the low-density models than in SCDM. This large effect, occurring at
relatively low redshifts of  $z\simeq$0.5 -- 1, provides a powerful method
for determining $\Omega _{m}$ and $\sigma _{8}$ from the observed abundance
of high-redshift massive clusters. The dependence of the cluster abundance
on $\Lambda $ is much weaker; it becomes more significant at higher
redshifts ($z\gtrsim $1) (e.g. Eke  et al. 1996, 1999, Bahcall et al. 1997, 
Fan  et al. 1997). 

The LCDM and QCDM mass functions evolve at roughly the same rate,
consistent with the prediction of \cite{WS98}.
OCDM evolves somewhat slower; the differences become more
pronounced at $z>1$.  
This can be quantified by considering the ratio of cluster
abundance $n(z=1)/n(z=0)$ for clusters of mass 
$\geq 8\times 10^{14}h^{-1}M_{\odot }$
(using the best-fit P-S approximation of $\S$\ref{secpscomp}).
OCDM with a ratio of $2.2\times 10^{-3}$ is roughly double
LCDM ($1.2\times 10^{-3}$) and QCDM ($1.1\times 10^{-3}$).
(SCDM is much lower at $6.9\times 10^{-8}$; see $\S$\ref{secobscomp}). 
With improved data sets, this will enable a discrimination
between flat and OCDM cosmologies (assuming $\Omega_{m}$ is known).
The evolution of the cluster MF
presented in Figure~\ref{figmfevol} can be 
used for direct comparisons with observations. 

The analysis presented above is for cluster masses within the standard Abell
radius of 1.5$h^{-1}$Mpc comoving. Many observations, but not all,
determine cluster masses within this radius (based on velocity dispersion,
gas temperature, or gravitational lensing). Some observations however
determine cluster masses within other radii (typically 0.5 or 1$h^{-1}$Mpc).
 In order to provide proper comparisons with observations, as well as
understand the dependence of the MF on the selected radius within which the
cluster mass is measured, we investigate the MF evolution for different
radii. In Figure~\ref{figmfrads} we present the evolution of the cluster 
MF for cluster
masses measured within radii of 0.5, 1, and 1.5$h^{-1}$Mpc, both comoving
and physical, for each of the four models. 

The dependence of the MF evolution on radius is evident in 
Figure~\ref{figmfrads}.  The
progression from large to small radii for either the comoving or physical
scales reflects mostly a progression from the large cluster mass within 
1.5$h^{-1}$Mpc to 
the smaller mass contained within 0.5$h^{-1}$Mpc 
for the \emph{same} clusters. 
Therefore, this progression simply shifts the MF to
smaller masses as the radius decreases (at all redshifts). 
If the cluster density profile is $\rho(R)\sim R^{-\gamma}$ on these scales, 
where $3-\gamma\approx 1$, then the mass inside $R$ is
$M(<R)\sim R^{3-\gamma}$ and
a similar shift of mass
with radius is expected--- as is indeed seen in the simulations.

A more interesting trend is seen between comoving and physical radii. The
evolution of the MF is much weaker for cluster masses determined within
physical radii than for comoving radii. The strong evolution seen for
comoving radii is greatly reduced for physical radii because the same mass
is contained within the larger physical radius (at high redshift) instead of
the smaller comoving radius (since R$_{\rm phy}$ = R$_{\rm com}(1+z)$). 
Therefore, for a fixed cluster mass (as a function of redshift) the 
use of physical
radius represents relatively lower mass clusters at high redshifts as
compared with the comoving radius, by a factor of approximately 
$(1+z)^{3-\gamma}$.
Since lower mass clusters are more
numerous, there will be more clusters at high redshifts for 
cluster masses defined within physical radii
than within comoving radii, 
thus yielding considerably weaker evolution for the former.
The effect is stronger in low density models, where only mild
evolution is seen in physical coordinates.  This is mostly
due to the fact that for a low density universe 
evolution is significant from early times until
$z\sim \Omega_{m}^{-1}\sim 3$, when the universe becomes
``open'' for local observers and both accretion and merging
are reduced.  In such models, clusters virialize at $z \sim 2$,
after which their evolution slows down; in fixed coordinates
their properties, including the mass function, approach constancy.

The difference is large; for example, the abundance of Coma-like clusters
($M_{1.5} \simeq 6\times 10^{14}h^{-1}M_{\odot }$) at $z=1$ is 
nearly 100
times larger for R = 1.5$h^{-1}$Mpc physical radius than for 1.5h$^{-1}$Mpc
comoving radius. 
It is therefore essential that observed 
and simulated clusters be compared using the same radii.
The results presented in Figure~\ref{figmfrads} can be used for comparisons
with observations of cluster masses within these different radii.

\section{Comparison with Press-Schechter Approximation} \label{secpscomp}

In this section we compare the simulation results with the
predictions of the Press \& Schechter (1974) 
approximation frequently used (e.g. Eke, Cole, \& Frenk 1996,
Fan, Bahcall, \& Cen 1997)
in the estimation of 
the number density and mass distribution of collapsed objects.
The P-S approximation states that, after smoothing on the appropriate length
scale, regions with density contrast above some threshold $\delta_c$
will have collapsed and virialized.  
The spherical collapse model
predicts $\delta_c=1.686$, with a weak dependence on $\Omega_m$
and $\Lambda$ \citep{ECF96, WS98}.  
Because P-S theory predicts the mass inside a virial radius, 
and this radius is often used in numerical work,
we first compare the standard P-S predictions with the simulated
mass function found using HOP.  Employing the HOP 
regrouping algorithm and tracing clusters to an outer overdensity  
threshold of 160 gives results similar to the 
Friends-of-Friends (FOF) algorithm with linking parameter 0.2,
which is frequently used to analyze numerical simulations
\citep{EH98, GBQTBKL99, JFWCCEY00}.  The resulting HOP mass function
is shown in Figure~\ref{figpshop}, along with the P-S prediction. 
The shape of the standard P-S mass function does not fit
the numerical results; it predicts too many low mass clusters 
and too few higher mass clusters.
This is now a well established drawback of the P-S
approximation \citep{GSPHK98,GBQTBKL99,JFWCCEY00}.  
Furthermore, at higher redshifts there are more collapsed
objects than predicted by standard P-S.

To explore the comparison in more detail,
and to allow for proper comparison with observations,
we will use the
mass function defined in terms of the observable quantity of mass
within a specific radius.
Because P-S theory predicts the mass inside a virial radius, an adjustment
needs to be made to predict the mass
within 1.5$h^{-1}$Mpc.  We follow the prescription of \cite{CMYE97}
relating the top-hat smoothing length to $M_{1.5}$ rather than
the virial mass.
A slope of $3-\gamma=0.6$ is assumed in the relation 
$M(<R)\propto R^{3-\gamma}$ near the virial radius. 
This slope is
consistent with the mass profile of observed rich clusters 
\citep{CYE97,RGDMW00}.
Also, since N-body simulations give
results in variance with P-S,
we treat $\delta_c$ as a free parameter.  For a given comoving
$M_{1.5}$, $\delta_c$ was adjusted so that the P-S prediction for 
$n(>M_{1.5})$ matched the N-body result at that mass.  
This was done for all models and redshifts, beginning at the mass
of the tenth most massive cluster and working down to 
$6\times10^{13}h^{-1}M_\odot$.

The resulting $\delta_c$ as a function of $M_{\rm 1.5com}$ is 
shown in Figure~\ref{figdeltac}.  It is apparent that the standard
P-S approximation
(using $\delta_c$ from spherical overdensity in linear theory)
does not fit the simulations.  
The cluster abundance depends on $\delta_c$ roughly as 
$\exp(-\delta_c^2/\sigma^2)$, so a higher $\delta_c$
means fewer clusters have formed.
The $\delta_c$ required by the simulation MF varies with mass.
In other words, the shape of the standard P-S mass function is
incorrect; for the low $\Omega_m$ models it predicts too many
low mass clusters (since the standard $\delta_c$ is not large
enough), even if
$\delta_c$ is fixed to match the high mass clusters
($\gtrsim 2\times10^{14}h^{-1}M_\odot$). 
The tilted SCDM model shows the opposite trend--- we are finding more
low mass clusters in the simulations than is predicted.  
This is due to the assumption that $M(<R)\propto R^{0.6}$ in adjusting
the virial mass to $M_{1.5}$.  For $\Omega_m=1$ the extrapolation
from $1.5h^{-1}$Mpc to $R_{vir}$ reaches to radii $<1h^{-1}$Mpc
for low mass clusters.
Allowing a steeper slope ($M(<R)\propto R$)
for smaller SCDM clusters with a virial radius less than 
1$h^{-1}$Mpc yields results similar to those seen in the other
models.

A second trend seen in Figure~\ref{figdeltac} 
is that $\delta_c$ is lower at
higher redshifts, showing that there are more collapsed
objects at $z>0$ than predicted by standard P-S.
The redshift dependence of $\delta_c$ can be parameterized as
$\delta_c=\delta_0+\delta_1/(1+z)$, with $\delta_0$ and $\delta_1$
chosen separately for each model.
A linear fit was made
to the best values at $M_{\rm 1.5com}=2\times10^{14}h^{-1}M_\odot$, and then
the parameters were adjusted slightly to reduce the difference between
the P-S predictions and simulations to below 10\% across the
entire mass range, if possible.  The final choices of $\delta_0$ and
$\delta_1$ are shown in Table~\ref{tbldeltac}.  It can be seen
that $\delta_0$ is close to the canonical value of 1.68
and that $\delta_1$ is $\lesssim 10$\% of this value,
meaning the $z$ dependence is weak.
Figure~\ref{fignofmcomp} shows both the N-body MF and the
results of the modified P-S formula using the $\delta_c$ 
from Table~\ref{tbldeltac}, along with the 
fractional difference between the two.  
The simulation data are fit well
by this modified P-S relation, to within 10\% for $z\lesssim 2$ 
(or $z\lesssim 1$ and $M>10^{14}h^{-1}M_\odot$ for SCDM).

An alternative to the simple power-law density profile used here
is the proposed universal profile for dark matter halos of
\citet{NFW97} (see also \citet{LM00});  the ``toy model'' of
\citet{BKSSKKPD00} provides the mass and redshift dependence of
the concentration.  
In order to test whether including this dependence 
would improve the analytic fit, the above analysis was repeated 
using the NFW profile to adjust the virial mass to $M_{1.5}$.
No improvement in the fit was seen;  $\delta_c$ shows
both mass and redshift dependence in a manner similar to that
observed using the power-law profile, and the fit is somewhat
worse for redshifts $z\gtrsim 2$.

Our results are in agreement with \cite{GBQTBKL99}, who likewise found 
the best fit $\delta_c$ to be slightly lower as redshift increases.  
However, the $z$ dependence found here is weaker. \cite{GBQTBKL99}
simulated an open model with parameters 
close to our OCDM run, and found $\delta_c\approx 1.65$ at $z$=1,
agreeing well with our value.  However, at $z$=0 the \cite{GBQTBKL99}
value is well above ours; this may be due to the fact that
their $\sigma_8$ value is higher as well.

\section{A Higher Resolution Run}  \label{secreseff}

An additional simulation was carried out using the same parameters
as the LCDM run, except with $1024^3$ particles,
allowing an examination of the effect of increasing the
numerical resolution.
The increased number of particles reduces the
mass of particle by a factor of eight to 
$7.75\times 10^{10}h^{-1}M_\odot$; the softening length was
chosen to be 13.6 $h^{-1}$kpc.  
The run took 800 PM steps and up to 9000 tree steps.
This simulation was carried out on a
256 processor SGI Origin 2000 at NCSA and took 9.5 days;
the considerable computational expense precludes further runs of this size.
For comparison,
the $N=1024^3$ Virgo ``Hubble Volume'' LCDM simulation 
\citep{JFWCCEY00,Evrard98} contained 27 Gpc$^3$, with
particle mass $2.25\times 10^{12}h^{-1}M_\odot$ and
Plummer softening length 100 $h^{-1}$kpc.

A comparison of the mass functions found in the $512^3$ and $1024^3$ 
runs is shown in Figure~\ref{figncomp}.  
It can be seen that for $z<2$ the results are 
similar, the main difference being that the higher resolution
run yields slightly more lower mass clusters:  at
$5\times10^{13}h^{-1}M_\odot$, there is a 10\% difference.
For typical rich clusters, with mass $\gtrsim 10^{14}h^{-1}M_\odot$,
the effect is negligible.
The modified Press-Schechter prediction, using $\delta_c$ as given in
Table~\ref{tbldeltac}, is also shown. 
At $z=0$, the simulation and modified P-S mass
functions are within 5\% of each other down to 
$n(>M_{1.5})=10^{-7}h^3$Mpc$^{-3}$  (that is, when there are
more than 100 clusters in the simulation volume).
At higher redshifts the modified P-S prediction yields 
roughly 10\%
fewer low mass clusters than 
the $1024^3$ simulation. 
Since increasing mass resolution by a factor
of eight and doubling force resolution has only a minor
effect, we feel confident the results presented here  will not be changed
significantly by increasing resolution.

\section{Comparison with Observations} \label{secobscomp}

The evolution of cluster abundance is compared with observations in
Figure~\ref{figmvsz}. Here 
we plot the abundance of massive clusters, above a given
mass threshold, as a function of redshift. We use the most distant
observed massive clusters since such clusters provide the most powerful test
of cosmological evolution: these massive clusters cannot exist at high
redshifts in $\Omega _{m}$ = 1 Gaussian models; they were simply not formed
yet (e.g., Peebles 1993, Fan  et al. 1997, Bahcall and Fan 1998,
Donahue  et al. 1999). The clusters we use are the same as those used
by Bahcall \&  Fan (1998): MS 0016+16, 0451-03, and 1054-03.
These clusters are at  $z=$0.55 and  $z=$0.83, 
with masses above a threshold of 
$M_{\rm 1.5com}\geq 8 \times 10^{14}h^{-1}M_{\odot }$. 
They have measured velocity dispersion 
($\sigma _{r}\gtrsim $ 1200 km/s), gas temperature (T $\gtrsim $ 8 Kev), 
strong S-Z decrements (Carlstrom 1998),
and (in 2 of the 3 clusters)
gravitational lensing mass determinations.
Also shown are clusters at
$z=0.38$ from the temperature function of \cite{Henry00}. The temperature
has been
converted to $M_{\rm 1.5com}$ using the \emph{observed} M-T relation: 
$M(\leq 1h^{-1}{\rm Mpc})=kT{\rm (keV)}\times 10^{14}h^{-1}M_{\odot }$,
with the minor extrapolation to $R_{\rm 1.5com}$ following the 
observed cluster mass profile (\cite{HOvK98},
Bahcall \& Fan 1998, Bahcall \& Sette 2000).  
The scatter in this observed relation is $\pm $ 25\%.

The model predictions are presented by the curves in 
Figure~\ref{figmvsz}.  The shaded
horizontal band indicates the range where 
only a single cluster is expected to be found in
the simulation box; below this band, where no clusters can be found, we
extrapolate the $n(z)$ curve with our modified P-S approximation 
($\S$\ref{secpscomp}) as
shown by the lighter curves. The comparison with observations is powerful:
no massive clusters are found in the $\Omega_{m}=1$ simulations at 
any redshifts $z>$ 0.2! 
A few clusters are found in the $\Omega _{m}$ = 0.3
simulations at  $z\sim 0.6$ and 0.8, consistent with observations. 
The fact that such massive clusters exist in the universe at  $z\simeq $ 0.8
and 0.6 rules out $\Omega _{m}$ = 1 Gaussian models at a very high
significance level: if $\Omega _{m}$ = 1, 10$^{-5}$ clusters would be
expected in the observed survey volume at  $z\simeq $ 0.8 while 1 cluster is
observed; 10$^{-3}$ clusters would be expected at  $z\simeq $ 0.6 while 2
are observed; and 4 x 10$^{-2}$ clusters expected at  $z\simeq $ 0.4 while 2
are observed. Equivalently, 
based on the observed density, 
the simulation box should contain 10 such
massive clusters at  $z\simeq $ 0.8 and 10 clusters at  $z\simeq $ 0.6,
while the $\Omega _{m}$ = 1 simulation
contains no clusters; from the modified P-S extrapolation one 
would expect only 10$^{-4}$ and 10$^{-2}$ clusters, respectively.
In the $\Omega _{m}$ = 0.3 simulations, on the
other hand, we find 1 cluster in the simulation box at $z=0.8$ and 3
clusters at $z=0.6$. An even better match with the 
observed cluster abundance would be
achieved with a somewhat lower $\Omega_{m}$ value of $\sim$0.2 (e.g.,
Bahcall \& Fan 1998). 
We note that
the observed cluster abundance at $z$=0.5--0.8 could be even higher 
than used above if the
observational surveys are incomplete; this would further reduce
the best fit value of $\Omega_{m}$.

The models with $\Omega_{m}$ $\simeq $ 0.3 (LCDM, QCDM, OCDM) provide a
good fit to the data. While a precise value of $\Omega _{m}$ needs to
await a larger and improved sample of massive high redshift clusters, the
current data--- the existence of just a few massive clusters at
high-redshift--- already shows that 
(for Gaussian models) the mass density of the universe is low:
$\Omega_{m}\sim 0.3$. 
These conclusions from direct large-scale simulations
reinforce previous estimates based on 
the standard P-S approximation (Bahcall  et al. 1997, 1998, Carlberg 
et al. 1997, Henry 1997, 1999, Eke et al. 1998, Donahue
et al. 1999, and references therein).

\section{Conclusions} \label{secconcl}

Gigaparsec scale high-resolution N-body simulations are used to
determine the cluster mass function from $z=3$ to $z=0$ for
four CDM models: LCDM, OCDM, QCDM, and tilted SCDM. These large,
high-resolution simulations, with 10$^{5}$ clusters of mass 
M$_{1.5} \geq 5\times 10^{13}h^{-1}M_{\odot }$ at the present day, 
allow an accurate determination of the MF evolution over this
redshift range.

The evolution of the  mass function is presented for cluster masses within 
several radii typically used in observations: 0.5, 1, and 1.5$h^{-1}$%
Mpc, both comoving and physical; this will enable direct comparisons with
different observations. The evolution of the MF
for clusters of a given mass within a physical radius is 
found to be considerably
weaker than for mass within a comoving radius; this is due to the 
relatively larger physical radii at high redshift, which means
that more numerous poorer clusters are included.
After virialization at $z \approx 2$ in the $\Omega_{m}=0.3$ models,
there is relatively little evolution in physical coordinates.
Because of the large differences in the resulting evolution, it is essential
that the cluster evolution be compared using 
the same simulated cluster mass--- within the same radius---
as observed.

We compare the simulation results with the Press-Schechter 
approximation; the latter predicts too many low mass clusters 
at $z=0$ and too few clusters at higher redshifts.
We
provide refined parameters for this approximation that best fit the
simulations as a function of redshift and mass.
With the chosen fits to 
the overdensity parameter $\delta_c$, the agreement with the simulation
MF is better than 10\% for $z\leq 2$ in the three $\Omega_m=0.3$ models
and $z\leq 1$ in SCDM.

We compare the simulation results with current observations of the
most massive distant clusters observed to  $z\simeq$0.8. We show that
Gaussian $\Omega _{m}$=1 models such as tilted SCDM, which predict
extremely rapid evolution from  $z\sim$1 to  $z\sim$0, are ruled
out at a high significance level; they produce no massive clusters at high
redshifts (10$^{-5}$ clusters per the observed volume), in contradiction
with observations. 
Showing much weaker evolution,
the $\Omega _{m}$ = 0.3 models--- LCDM, OCDM, 
and QCDM--- are all 
consistent with the data. These results provide one of the most
powerful constraints on the mass density of the universe.
Upcoming surveys of distant clusters should be able to provide
further discrimination between flat (LCDM or QCDM) and open
(OCDM) low density cosmologies, since the latter shows
relatively weaker evolution.

%%%%%% Acknowledgments %%%%%% 
\noindent
This research was supported by NSF Grants AST-9318185 and
AST-9803137 (under Subgrant 99-184), and the NCSA Grand Challenge 
Computational Cosmology Partnership under NSF
Cooperative Agreement ACI-9619019, PACI Subaward 766.

% No more than seven \figcaption commands are allowed per page

%\clearpage

\plotone{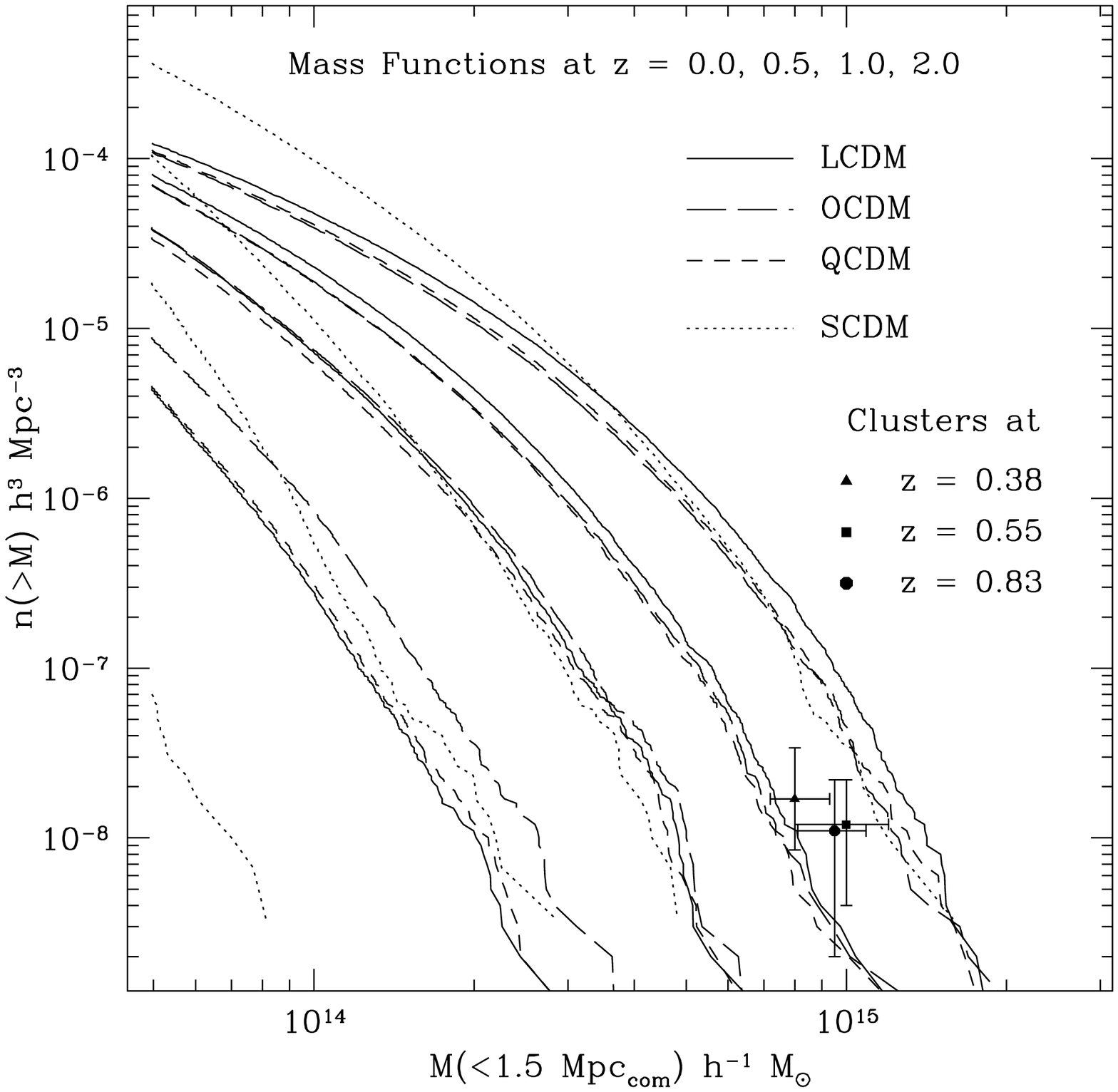}
\figcaption[f1.eps]{The cluster mass functions at 
redshifts of $z =$ 0, 0.5, 1.0, and 2.0 for the 
four different cosmological models are plotted as lines.  Note that the
SCDM model (dotted line) at $z=0.5$ is comparable to the low density 
models at $z=1$
and the SCDM model at $z =$ 1 is comparable to the 
low density models at $z=2$.
The three 
data points with error bars are from observational data discussed 
in $\S$\ref{secobscomp}.
\label{figmfevol} }

\epsscale{0.58}
%\plotone{figmfrads.eps} 
\plotone{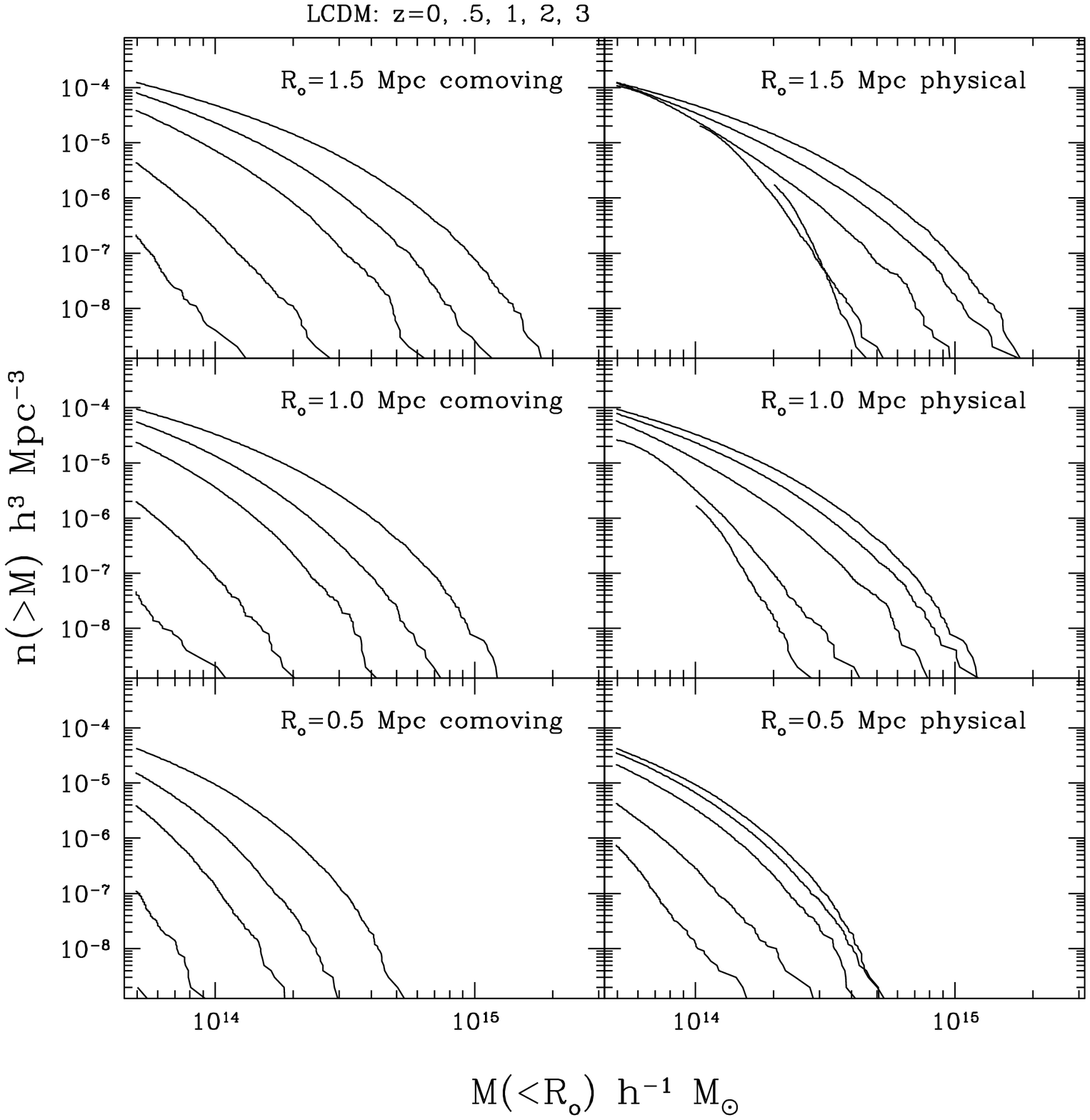} 
\plotone{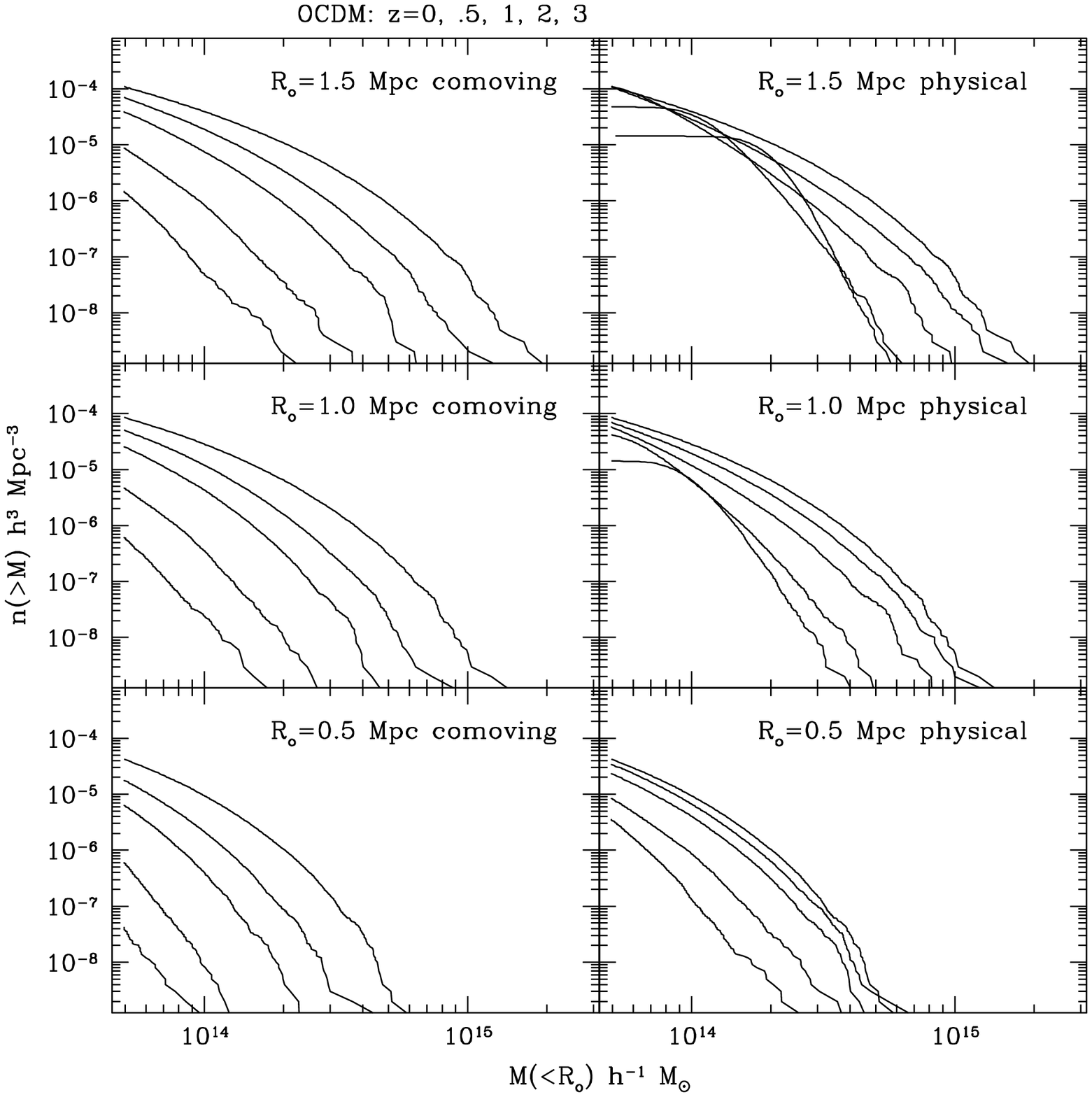} 
\plotone{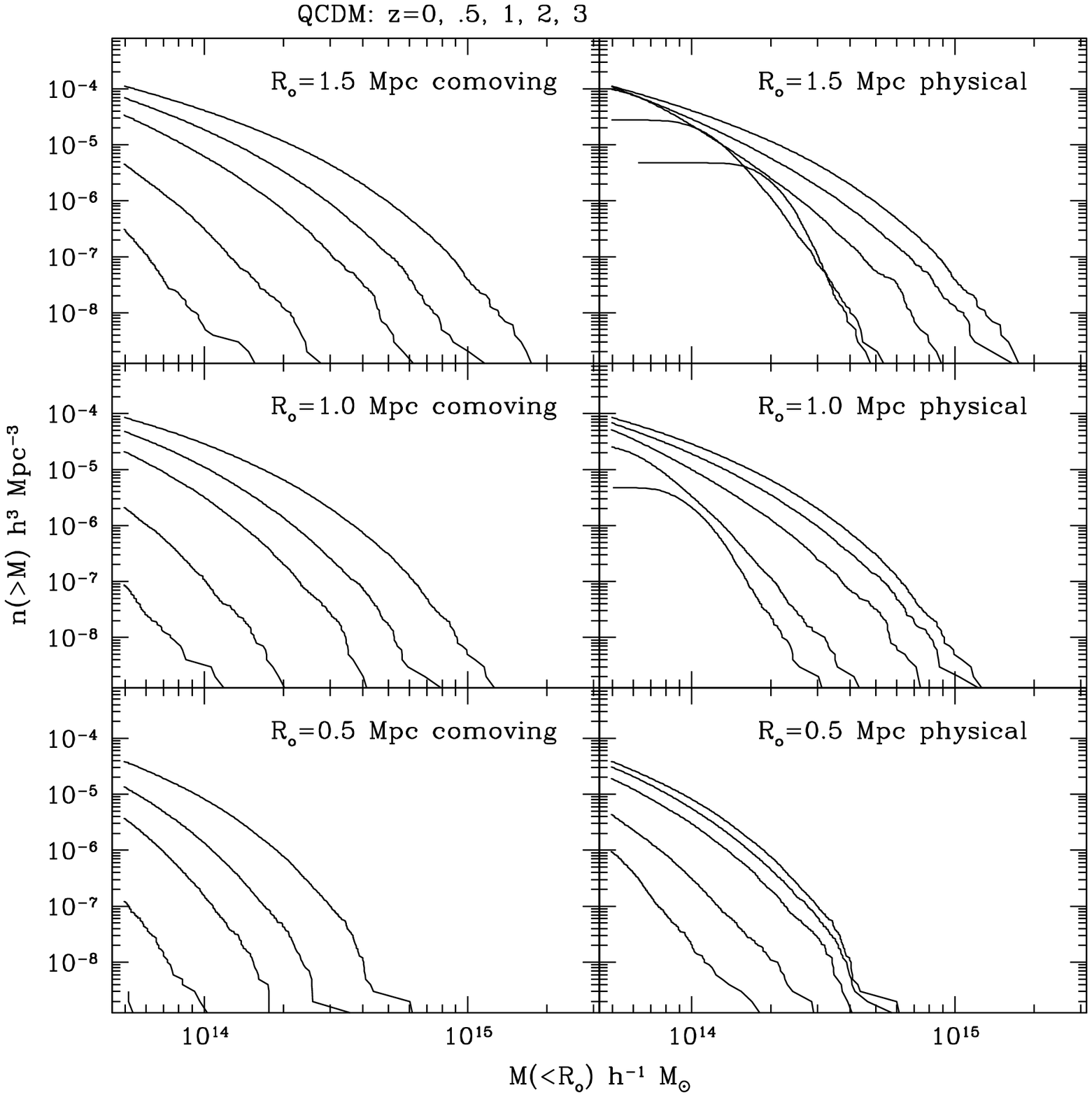} 
\plotone{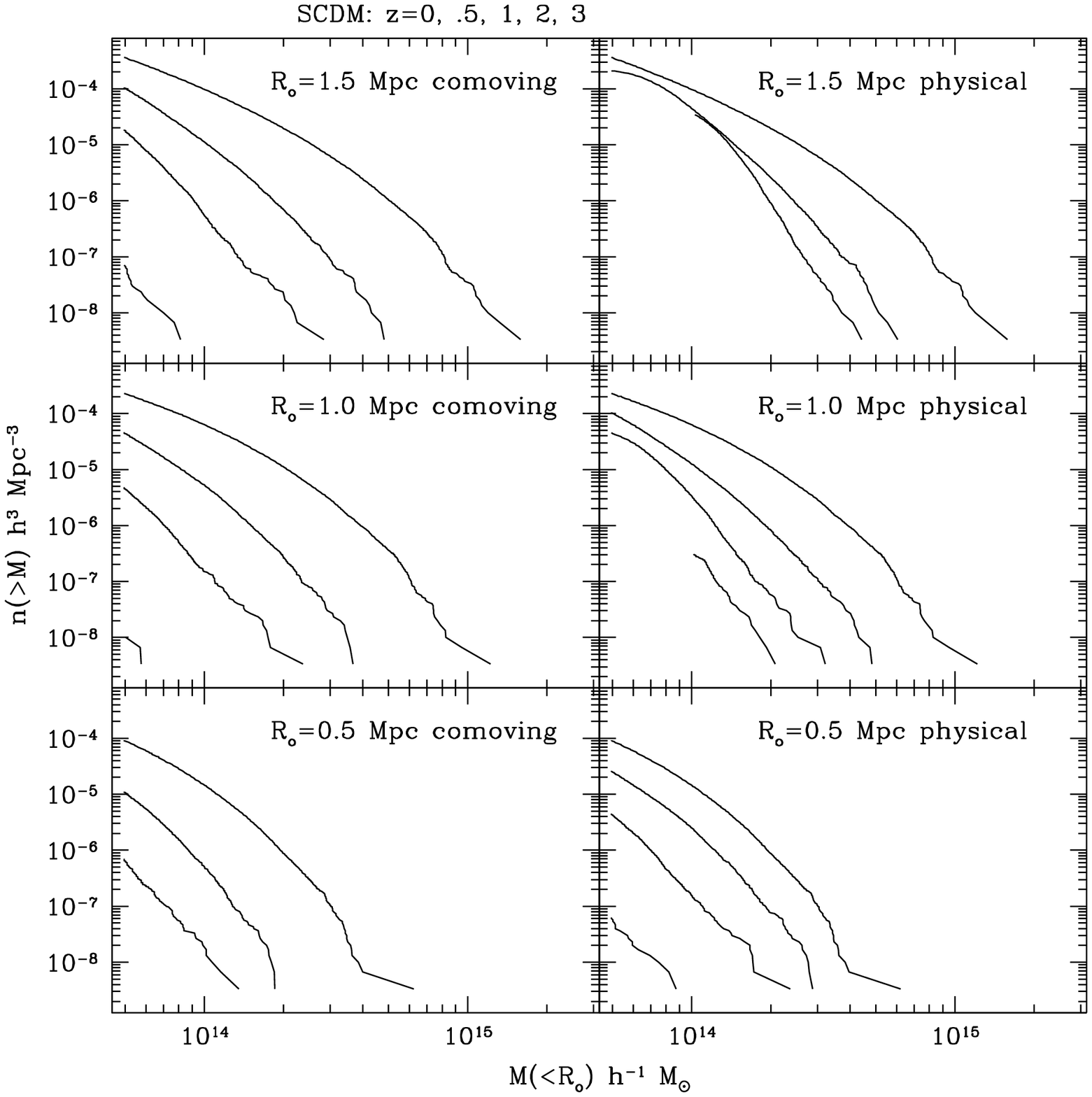} 
\epsscale{1.0}
\figcaption[figmfrads.eps]{The evolution of the cluster MF for cluster
masses measured within radii of 0.5, 1, and 1.5h$^{-1}$Mpc, both
comoving and physical, for each of the four models.
\label{figmfrads} }

\epsscale{0.8}
\plotone{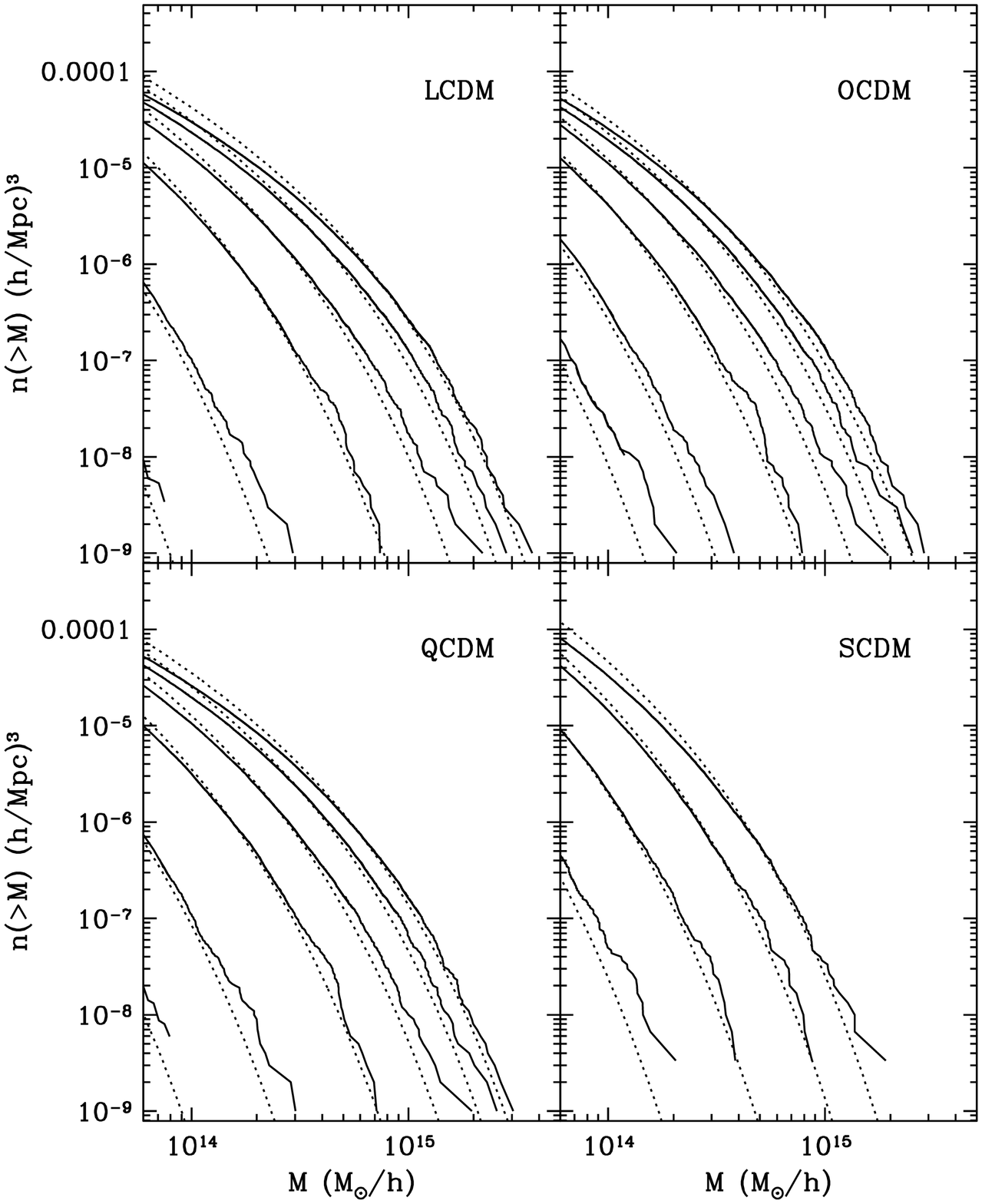}
\epsscale{1.0}
\figcaption[f3.eps]{
Solid lines: the simulated mass function, using masses determined
via the HOP algorithm.
Dotted lines: Press-Schechter approximation, with $\delta_c$
from linear theory. 
The curves from top to bottom are for redshifts $z=0, 0.17, 0.5, 1, 2, 3$.
\label{figpshop} }

\plotone{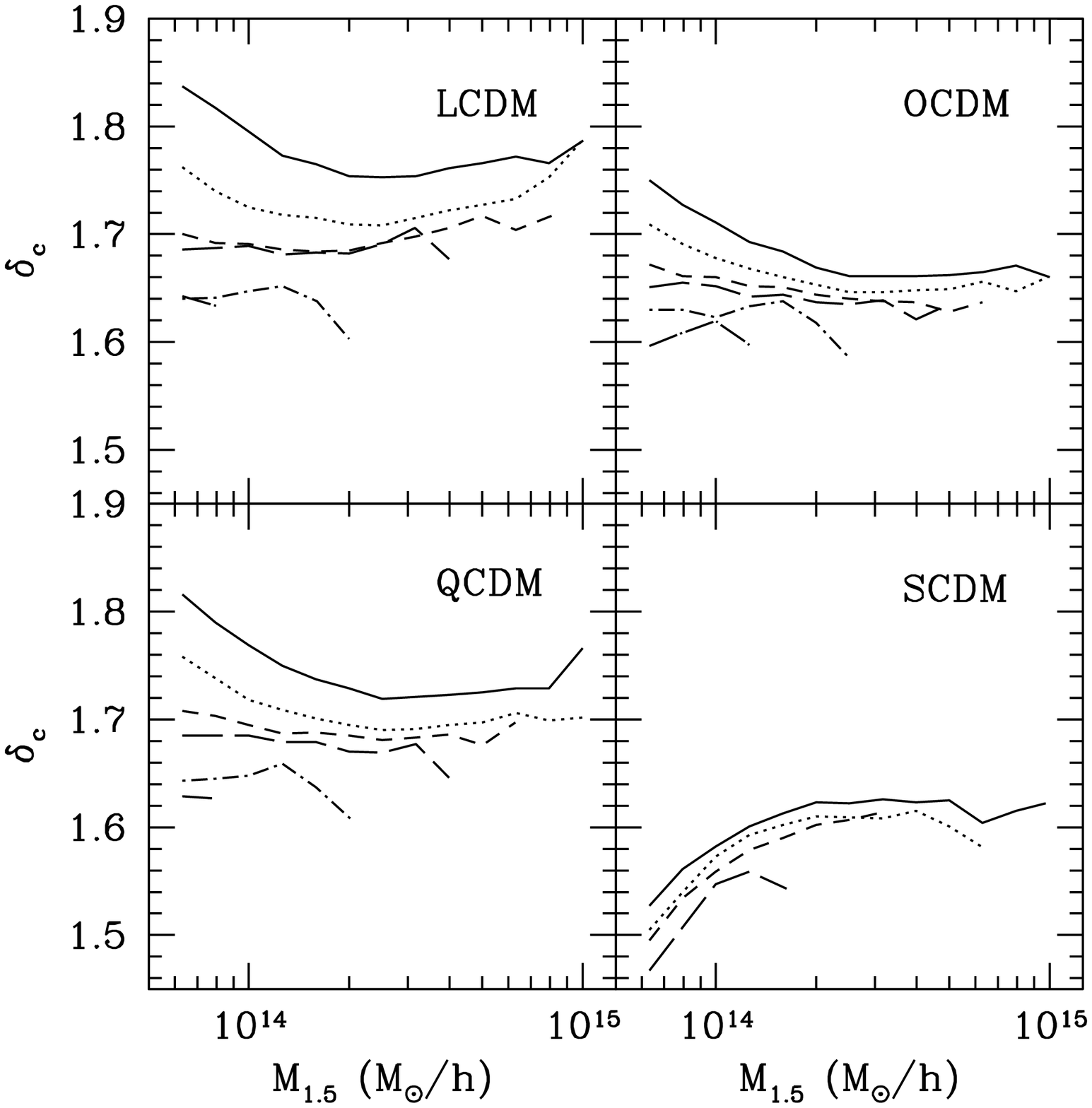}
\figcaption[f4.eps]{The best fit P-S overdensity parameter $\delta_c$
as a function of mass. The curves from top to bottom are for redshifts
$z=0, 0.17, 0.5, 1, 2, 3$.  Virial mass is translated to mass within
comoving 1.5$h^{-1}$Mpc assuming the observed mass profile
$M(<R)\propto R^{0.6}$.
\label{figdeltac} }

\epsscale{0.8}
\plotone{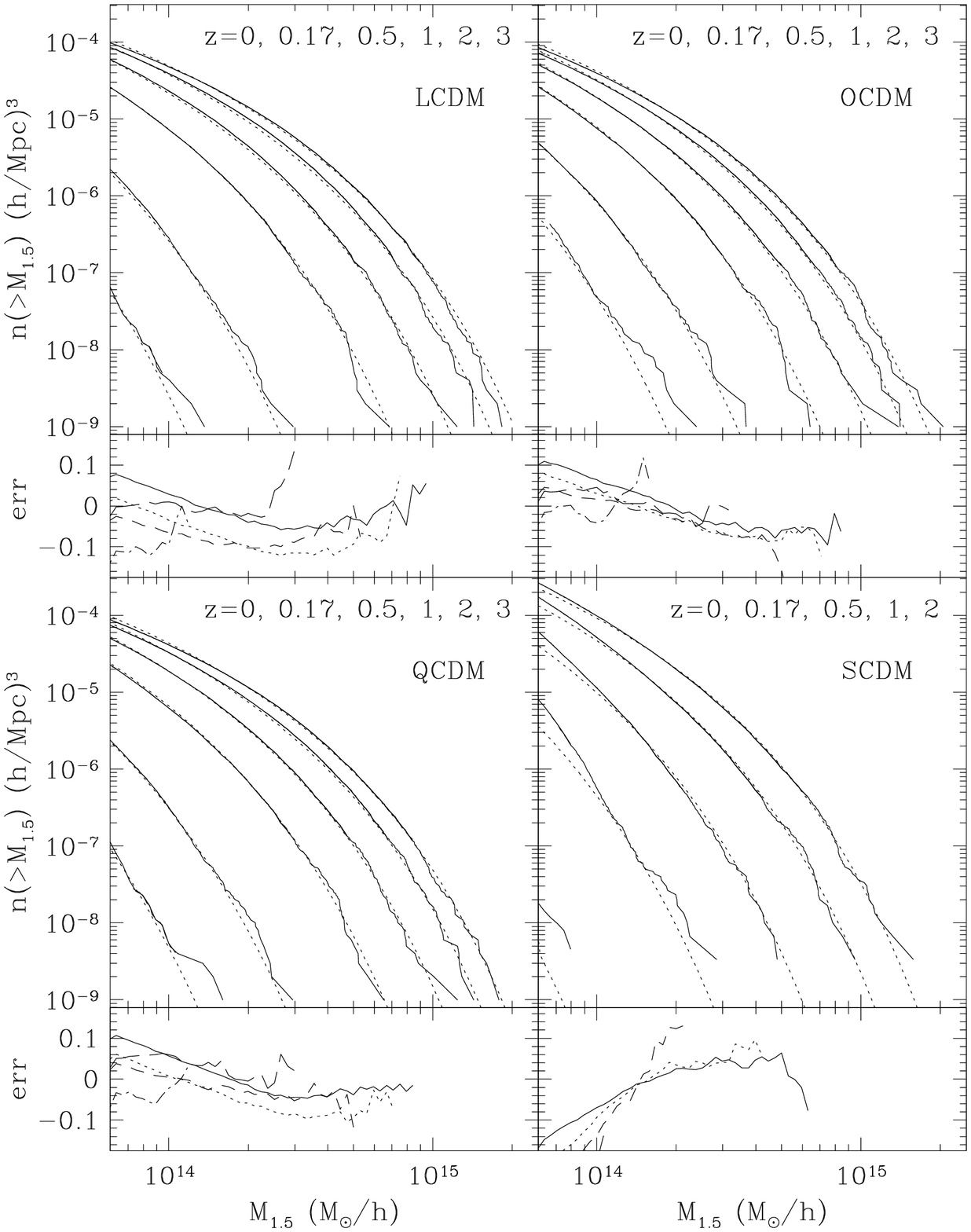}
\epsscale{1.0}
\figcaption[f5.eps]{The larger panels show the number density
as a function of mass from the simulations (solid line) and from 
the modified P-S
approximation (dotted lines) using the overdensity parameter from 
Table~\ref{tbldeltac}.  
The smaller panel beneath each of these plots shows the fractional 
difference between the two, n(PS)/n(sim)-1.
The line types match those in Figure~\ref{figdeltac}.
\label{fignofmcomp} }

\plotone{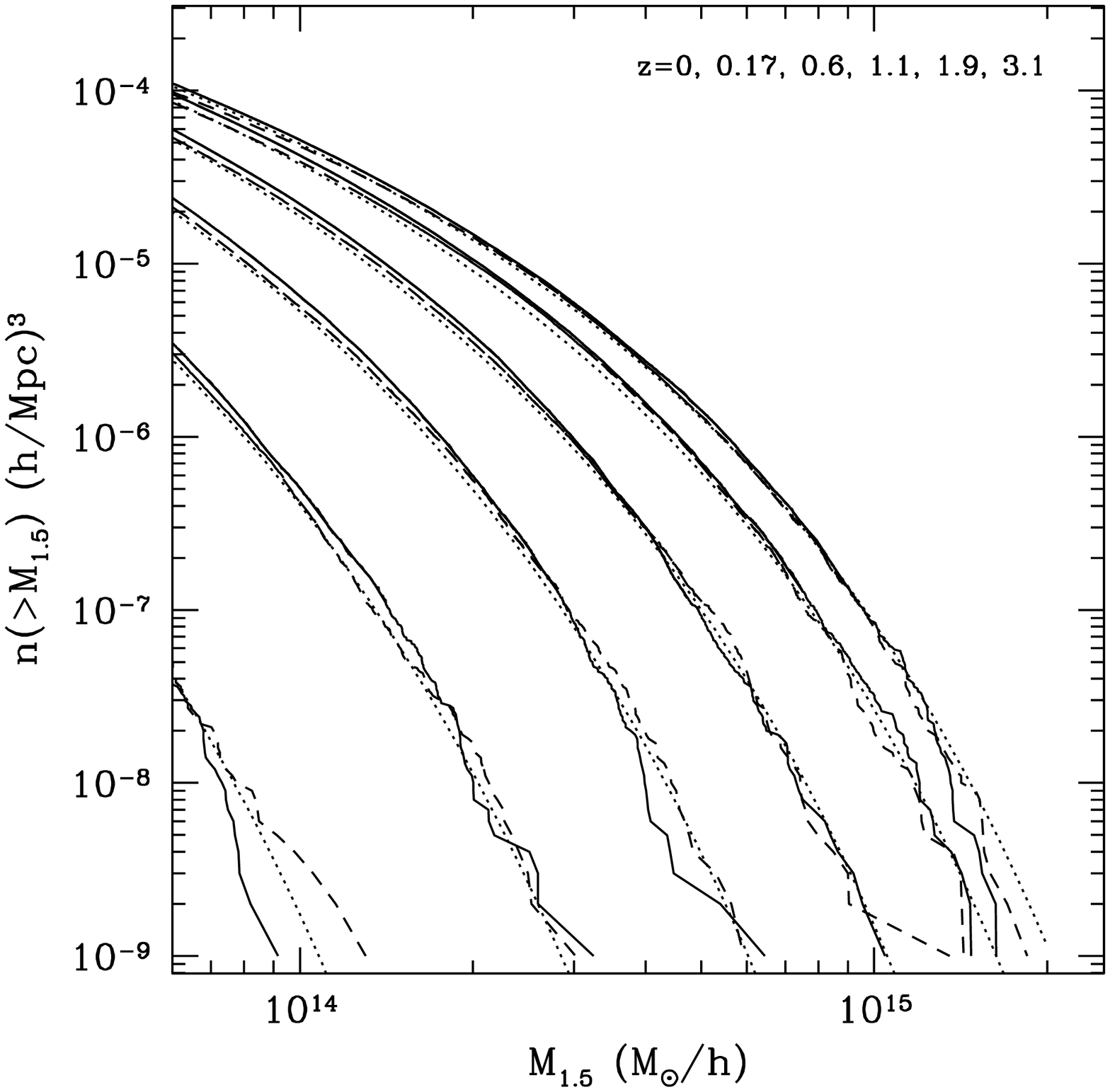}
\figcaption[f6.eps]{A comparison of the mass function in the 
LCDM model using $512^3$ (dashed lines) and $1024^3$ (solid lines) 
particles.  Also shown is the modified Press-Schechter prediction 
(dotted lines), using $\delta_c$ as given by Table~\ref{tbldeltac}.
\label{figncomp} }

\plotone{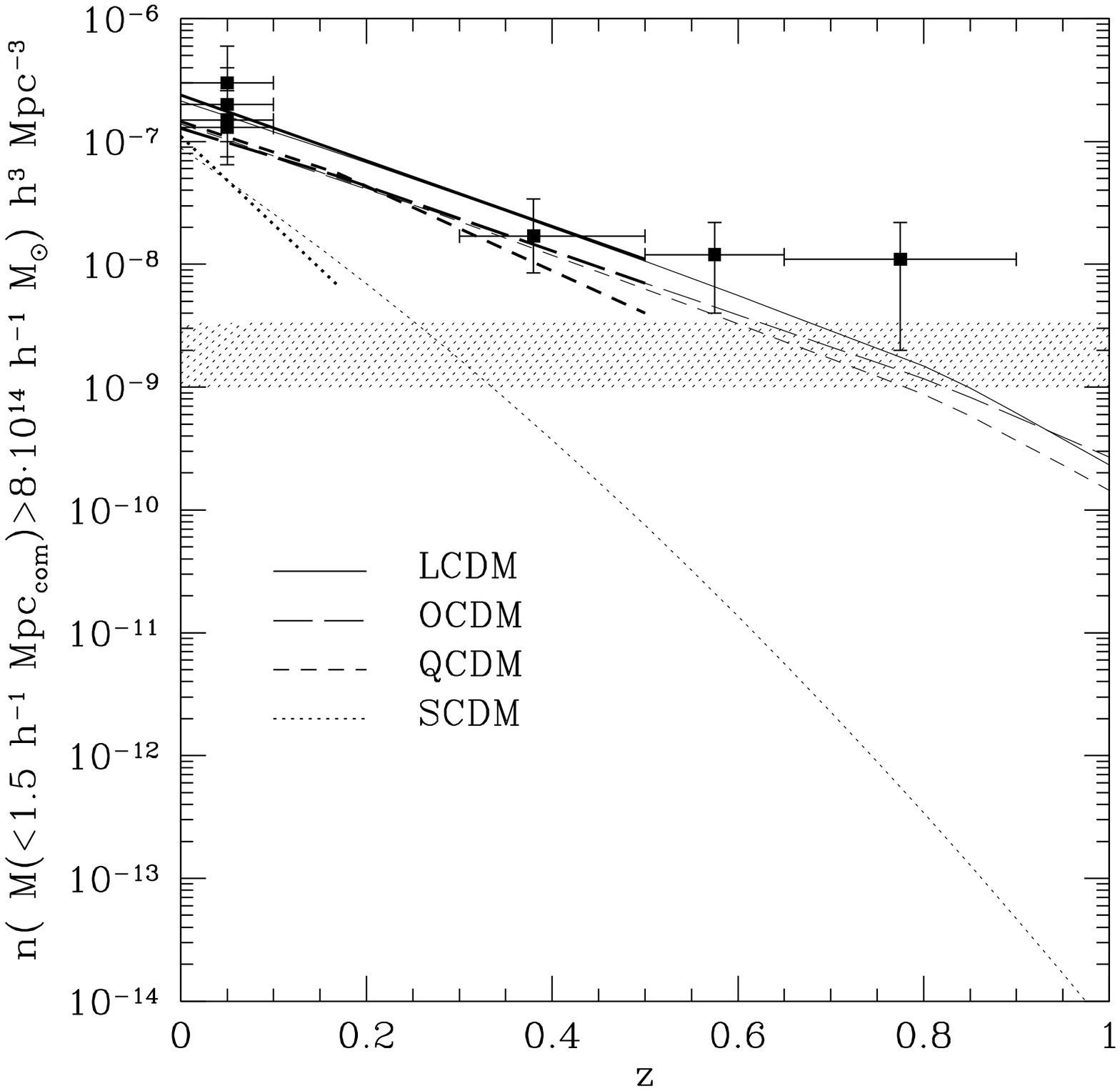}
\figcaption[f7.eps]{The abundance of massive clusters, above a 
given mass threshold, as a function of redshift. The dark lines are from
our numerical simulations, while the light lines are 
the modified P-S approximations using
the parameters from 
Table~\ref{tbldeltac}.  The data points are from observations discussed
in $\S$\ref{secobscomp}.  The dotted region indicates the range of densities for
which the numerical simulations would only have one cluster in the entire
volume.
\label{figmvsz} }

\clearpage
\begin{deluxetable}{lrrrrrlrl}
\tablecolumns{9}
\tablecaption{Model Parameters \label{tblparams} }
\tablewidth{0pt}
\tablehead{ 	\colhead{Model}	& 
		\colhead{$\om$}	&
		\colhead{$\ov$}	&
		\colhead{$\ob$}	&
		\colhead{$\oc$}	&
		\colhead{$h_{100}$}	&
		\colhead{n}	&
		\colhead{$\sigma_8$}	&
		\colhead{L\tablenotemark{a}} }
\startdata
LCDM & 0.30 & 0.70 & 0.040 & 0.260 & 0.67 & 1.0   & 0.90 & 1000 \\
OCDM & 0.30 & 0.00 & 0.040 & 0.260 & 0.67 & 1.0   & 0.80 & 1000 \\
QCDM & 0.30 & 0.70\tablenotemark{b} 
		   & 0.040 & 0.260 & 0.67 & 1.0   & 0.84 & 1000 \\
SCDM & 1.00 & 0.00 & 0.076 & 0.924 & 0.50 & 0.625 & 0.50 & 669.4
\enddata
\tablenotetext{a}{box length in $h^{-1}$Mpc}
\tablenotetext{b}{actually a Q component with w=-2/3}
\end{deluxetable}

\begin{deluxetable}{lcc}
\tablecolumns{3}
\tablecaption{Best-fit $\delta_c=\delta_0+\delta_1/(1+z)$ \label{tbldeltac} }
\tablewidth{0pt}
\tablehead{\colhead{Model} & \colhead{$\delta_0$} & \colhead{$\delta_1$} }
\startdata
LCDM &  1.60 & 0.18 \\
OCDM &  1.61 & 0.07 \\
QCDM &  1.61 & 0.13 \\
SCDM &  1.53 & 0.09
\enddata
\end{deluxetable}

\end{document}